\documentclass
[aps,prl,amsmath,amsfonts,amssymb,twocolumn,superscriptaddress,showpacs]{revtex4}%
\usepackage{times}
\usepackage{amsfonts}
\usepackage{amsmath}
\usepackage{amssymb}
\usepackage{graphicx}%
\providecommand{\U}[1]{\protect\rule{.1in}{.1in}}

\begin{document}

\title{Locally Inaccessible Information as a Fundamental Ingredient to Quantum Information}

\author{F. F. Fanchini}
 \email{fanchini@iceb.ufop.br}
\affiliation{Departamento de F\'isica, Universidade Federal de Ouro Preto, CEP 35400-000, Ouro Preto, MG,
Brazil}
\author{L. K. Castelano}
\affiliation{Departamento de F\'isica, Universidade Federal de S\~ao Carlos, CEP 13565-905, S\~ao Carlos, SP, Brazil}
\author{M. F. Cornelio}
\affiliation{Instituto de F\'{\i}sica Gleb Wataghin, Universidade Estadual de Campinas, CEP 13083-859,
Campinas, SP, Brazil}
\author{M. C. de Oliveira}
 \email{marcos@ifi.unicamp.br}
\affiliation{Instituto de F\'{\i}sica Gleb Wataghin, Universidade Estadual de Campinas, CEP 13083-859,
Campinas, SP, Brazil}
\affiliation{Institute for Quantum Information Science, University of Calgary, Alberta T2N 1N4, Canada}

\begin{abstract}
Quantum discord (QD) measures the fraction of the pairwise mutual 
information that is locally inaccessible, in a multipartite system. 
Fundamental aspects related to two important measures in quantum 
information theory the Entanglement of Formation (EOF) and the 
conditional entropy, can be understood in terms of the distribution of 
this form of Local Inaccessible Information (LII). As such, the EOF for 
an arbitrarily mixed bipartite system $AB$ can be related to the gain or 
loss of LII due to the extra knowledge that a purifying ancillary system 
$E$ has on the pair $AB$. Similarly, a clear meaning of the negativity 
of the conditional entropy for $AB$ is given. We exemplify by showing 
that these relations elucidate important and yet not well understood 
quantum features, such as the bipartite entanglement sudden death and 
the distinction between EOF and QD for quantifying quantum correlation. 
For that we introduce the concept of LII flow which quantifies the LII 
shared in multipartite system when a sequential local measurements are 
performed.\end{abstract}

\date{\today}
 \maketitle
\section{Introduction}
Different ways to measure quantum correlations have been widely studied in the last years \cite{qd, henderson, measures}. Among these quantum correlations, quantum discord \cite{qd} has played an important role. Based on the difference of two distinct definitions of the mutual information, Ollivier and Zurek developed a new measure of quantum correlations. This new feature of correlations was explored in its various aspects \cite{aspects,maziero}, intriguing the community by its peculiar properties - for instance, asymmetry and sudden changes \cite{schange}. It was recently shown that the Entanglement of Formation (EOF) and Quantum Discord (QD) obey a very special monogamic relation \cite{power}. This important result gives rise to new operational aspects for quantum discord, such as the net amount of entanglement processed in a quantum computer \cite{power}, as the difference between the entanglement cost and entanglement distillation \cite{marcio}, and as the amount of entanglement consumed in the state merging protocol \cite{cavalcanti}.

Differently from classical systems, a fraction of the quantum mutual information can not be accessed locally. Based on this idea, other interesting operational interpretation of QD emerges - as a measure of the mutual information fraction that is not accessible locally or, shortly, the locally inaccessible information (LII) \cite{demon}.
In this paper, we explore the properties of the LII to derive fundamental relations - We show that EOF between any two subsystems $A$ and $B$ can always be written exclusively as a function of the LII. Moreover it is possible to write the EOF between two subsystems $A$ and $B$ as average LII of the pair minus the balance of LII of the pair with a purifying environment $E$, giving to EOF a new operational meaning. We derive several relations between EOF and symmetrized and antisymmetrized versions of the LII that essentially quantify the average of the LII and the directional balance of LII, when measurements are made at $A$ and $B$, respectively. This allows for example to understand the difference between EOF and  QD for a bipartite system and elucidates important aspects of the entanglement sudden death. Furthermore, we relate the QD with the conditional entropy in a simple manner for an arbitrary bipartite system. Such a relation gives a new way to understand the negative signal of the conditional entropy.

\section{Locally inaccessible information}
In classical information theory, the mutual information (MI)  measures the amount of correlation between two stochastic variables, as measured by the Shannon entropy. The same concept when extended to quantum systems, in terms of the von Neumann entropy, allows the interpretation of the MI as the quantity of information shared by two quantum systems. It is generally accepted as the measure of the total amount of correlations (quantum and classical) of a quantum state. For a bipartite state $\rho_{AB}$, the quantum MI $I_{AB}$ accepts the extension of the standard form of the classical mutual information as
\begin{equation}
I_{AB} = S_A + S_B - S_{AB},\label{Iq}
\end{equation}
where $S_{AB}\equiv S(\rho_{AB})$, $S_{A}\equiv S( Tr_B\{\rho_{AB}\})$, and $S_{B}\equiv S( Tr_A\{\rho_{AB}\})$, where $S(\cdot)$ denotes the von Neumann entropy. However the very definition of the MI, $S(A:B)=S_A-S_{A|B}$ in terms of the conditional entropy $S_{A|B}=S_{AB}-S_{B}$ shows that there may be a problem with this simple extension above. In fact this second definition of the quantum MI is measurement-dependent and so, dependent on which system the measurement is performed. Thus from the start it seems that $S(A:B)$ is not necessarily symmetric, i.e., generally $S(A:B)\neq S(B:A)$. Moreover local measurement over a subsystem depends on the basis of the meter, and even with a good
basis choice, generally the total mutual information can not be accessed. Therefore a fraction of this
mutual information is non-local, the so-called local inaccessible information.

\begin{figure}[tbp]
\begin{center}
\includegraphics[width=.35\textwidth]{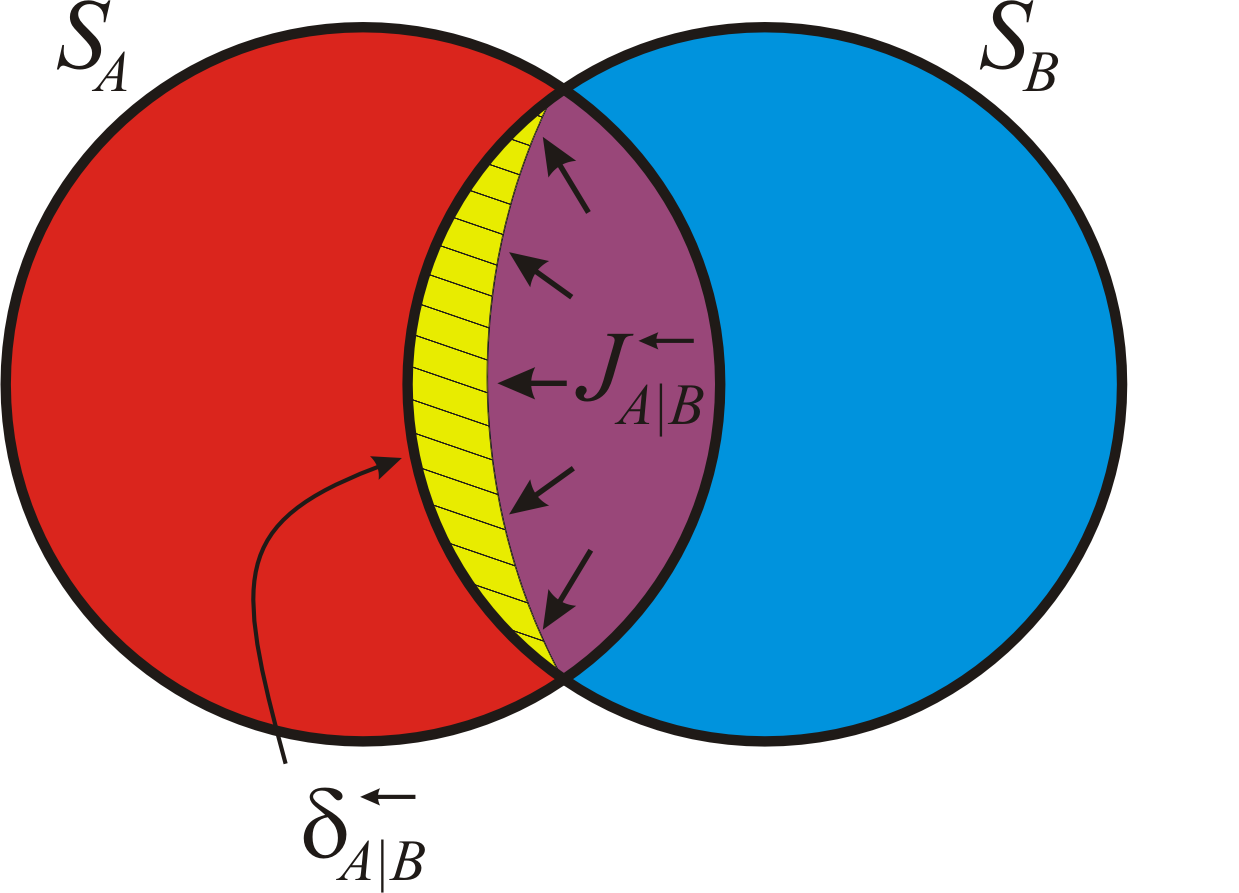} {}
\end{center}
\caption{(Color Online) An extended Venn diagram where the quantum entropies are exposed. Here a part of the mutual information is not locally accessible and it is divided in two parts: the classical correlation and the quantum discord.}\label{fig001}
\end{figure}

Given this peculiarity of a quantum system, Henderson and Vedral \cite{henderson} and, independently, Ollivier and Zurek \cite{qd} defined a quantity that measures the maximum amount of locally accessible information \cite{henderson},
\begin{equation}
J_{AB}^\leftarrow =\max_{\left\{ \Pi
_{k}\right\} }\left[S_A-\sum_{k}p_{k}S_{A|k}\right],\label{2}
\end{equation}
where $S_{A|k}$ is the conditional entropy after a measurement in $B$.
Explicitly, $S_{A|k}\equiv S(\rho_{A|k})$ where $\rho_{A|k}=\mathrm{Tr}_{B}(\Pi _{k}\rho_{AB} \Pi _{k})/\mathrm{Tr}_{AB}(\Pi_{k}\rho_{AB}\Pi_{k})$ is the reduced state of $A$
after obtaining the outcome $k$ in $B$ and $\{\Pi_k\}$ is a complete set of positive operator valued measurement that results in the outcome $k$ with probability $p_k=\mathrm{Tr}_{AB}(\Pi_{k}\rho_{AB}\Pi_{k})$. In this case, since a measurement might give different results depending on the basis choice, a maximization is required.
Thus $J_{AB}^\leftarrow$ is the locally accessible mutual information
and gives the maximum amount of $AB$ mutual information that one can extract by measuring at $B$ only \cite{demon}.
An illustration of that is shown in Fig. (\ref{fig001}), where the arrows represent the maximization involved in the calculation of the locally accessible mutual information. Note that a fraction of the MI is not locally accessible because it can be divided in two terms: one given by the $J_{AB}^\leftarrow$ and another given by the LII. The LII is then given by the MI minus $J_{AB}^\leftarrow$, which is exactly the definition of the QD,
\begin{equation}
\delta_{AB}^\leftarrow = I_{AB} - J_{AB}^\leftarrow.\label{discordia}
\end{equation}
In other words, the QD above gives the amount of information that is not accessible locally by measurements on $B$.
It is easy to see that the $\delta_{AB}^\leftarrow$, in fact, measures the difference between the conditional entropy given by the second term of Eq.~(\ref{2}), $S_q(A|B)\equiv\min_{\left\{ \Pi
_{k}\right\} } \sum_{k}p_{k}S_{A|k}$, under optimal measurements \cite{footnote1} on $B$ and the conditional entropy $S_{A|B}= S(A,B)-S(B)$ prior measurement, \begin{equation}
\delta_{AB}^\leftarrow =S_q(A|B) - S(A|B).
\end{equation} If  $S_q(A|B)=S(A|B)$ all the available information about $\rho_{AB}$ was acquired locally. So the QD has a strikingly simple meaning as a measure of how much a bipartite system state is affected by local measurements.
In fact,  the QD, $\delta_{AB}^\leftarrow $, vanishes
{
if and only if the density matrix of the composed system $\rho_{AB}$ remains unaffected by a measurement in $B$.}
In this case, all the MI between the pair is locally accessible. Based on this fact, we can rephrase the definition of $\delta_{AB}^\leftarrow$ as \textit{the fraction of the $AB$ Mutual Information Locally Inaccessible by $B$}.

While in $\delta_{AB}^\leftarrow$ the measurements over the basis that minimizes
the inaccessible information
are made over B (meaning that it is the mutual information of $AB$ that is inaccessible by $B$, which is being minimized),
in  $\delta_{BA}^\leftarrow$ those measurements are made over $A$ (meaning that the mutual information of $AB$ is inaccessible by $A$).
Indeed, there are states such that $\delta_{BA}^\leftarrow\neq 0$ though  $\delta_{AB}^\leftarrow=0$ and vice versa.
By using the asymmetry of $\delta_{AB}^\leftarrow$ and $\delta_{BA}^\leftarrow$, we can define two important quantities: The first one is the \textit{average} of the LII when measurements are made on $A$ and $B$,
\begin{equation}
\varpi^+_{A|B} = \frac{1}{2}\left(\delta_{AB}^\leftarrow + \delta_{BA}^\leftarrow\right),\label{polp}
\end{equation}
and the second one is the \textit{balance} of LII when measurements are made on $A$ and $B$,
\begin{equation}
\varpi^-_{A|B} = \frac{1}{2}\left(\delta_{AB}^\leftarrow - \delta_{BA}^\leftarrow\right).\label{polm}
\end{equation}
The \textit{average} LII  (Eq. (\ref{polp})) is a symmetric function since $\varpi^+_{A|B}=\varpi^+_{B|A}$ and quantifies how much a system state is disturbed by \textit{any} local measurement. On the other hand, the LII \textit{balance} \cite{demon} is asymmetric and gives the difference in the efficiency that each subsystem has to determine the mutual information by local measurements, which in sense quantifies the asymmetry of a given bipartite state under local measurements.
Suppose, for example, that $\varpi_{A|B}^->0$.
{In this case, a well chosen measurement in $A$ is more efficient for inferring mutual information of $AB$ than a well chosen measurement in $B$.
Thus, $A$ has less LII than $B$ and this imbalance increases as $\varpi^-_{A|B}$ increases.
On the other hand, if $\varpi_{A|B}^-<0$, then measurements in $A$ are less efficient for inferring the state of $B$ than vice versa.}
As seen below, these quantities are very useful to uniquely relate EOF to LII.

To present the relation between EOF and the LII, we begin by considering a pure joint state $|\psi_{AB}\rangle$. In this case, QD is symmetric ($\delta_{AB}^\leftarrow = \delta_{BA}^\leftarrow$) and is equal to EOF. Thus we can write
\begin{equation}
E_{AB} = \varpi^+_{A|B},\label{pure}
\end{equation}
where $\varpi^+_{A|B}$ is given by Eq.~(\ref{polp}), and so for an arbitrary pure bipartite state the EOF is simply the average LII.
Now we extend our consideration for an arbitrary mixed state $\rho_{AB}$ shared by $A$ and $B$. In such a case, a new subsystem $E$ that purifies the pair $A$ and $B$ must be considered.
In this new situation, an informational cost must be paid to include an additional subsystem - the exceeded knowledge that the environment $E$ has over the pair needs to be considered. As seen bellow, the EOF for the resulting mixed state $\rho_{AB}$ cannot be simply written as in Eqs. (\ref{pure}). Instead, it is given by the average LII of the pair ($A,B$) minus the LII balance of each of the subsystems $A$ and $B$ with $E$. To prove this relationship, let us suppose a pure state described by $\rho_{ABE}=|\phi_{ABE}\rangle\langle\phi_{ABE}|$ where $\rho_{AB}=Tr_E\{\rho_{ABE}\}$.
We begin with a conservation relation for the distributed EOF and QD derived earlier \cite{power},
\begin{eqnarray}
E_{AB} + E_{AE} &=& \delta^{\leftarrow}_{AB} + \delta^{\leftarrow}_{AE},\label{law1}\\
E_{AB} + E_{BE}&=& \delta^{\leftarrow}_{BA} + \delta^{\leftarrow}_{BE},\label{law2}\\
E_{AE} + E_{BE} &=& \delta^{\leftarrow}_{EA} + \delta^{\leftarrow}_{EB}.\label{law3}
\end{eqnarray}
Rearranging Eqs.(\ref{law1}-\ref{law3}) and writing them out in function of the average LII, given by Eq. (\ref{polp}), and the LII balance (Eq. (\ref{polm})), we can rewrite $E_{AB}$ as
\begin{equation}
E_{AB}=\varpi^+_{A|B}-\varpi^-_{E|A}-\varpi^-_{E|B}.
\label{Eab2}
\end{equation}
We can see that when compared to the pure state version from Eq. (\ref{pure}) the EOF in Eq. (\ref{Eab2}) decreases if the local measurements at the ancilla $E$ has less access to the mutual information with $A$ and $B$ than the subsystems $A$ and $B$ together. So the EOF is not  only given by the shared non-local information as in Eq. (\ref{pure}), but as well by the balance of the bipartite system $AB$ LII with the ancilla $E$. This relation allows an alternative interpretation of the he EOF, which is independent on the number of system copies \cite{bennett96} - \textit{ The EOF $E_{AB}$, for an arbitrarily mixed quantum state $\rho_{AB}$ is the average LII of $A,B$  minus the LII balance between each subsystem $A$ and $B$ with a purifying ancilla $E$}.  In simple words, the EOF of the pair $A, B$ is their average LII minus the loss (or gain) of LII due to correlation with $E$.


\begin{figure}[tbp]
\begin{center}
\includegraphics[width=.4\textwidth]{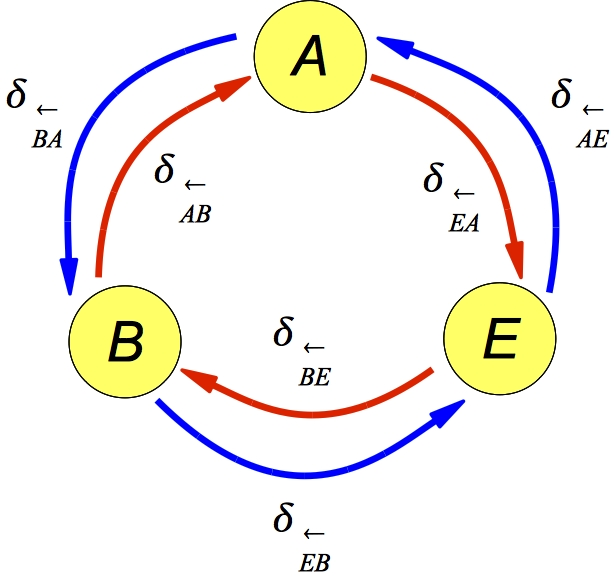} {}
\end{center}
\caption{(Color Online) Depiction of Clockwise (red arrows) and Counterclockwise (blue arrows) flow of Locally Inaccessible Information. The sum of the two possible directions of LII Flow results in the sum of all possible EOF between pairs $A$, $B$ and $E$. }\label{flow}
\end{figure}
We shall return to discuss the implications of Eq. (\ref{Eab2}) soon, but first we must define another way to interpret how the LII is distributed in the system.

\section{entanglement of formation and the flow of lii}
Since the quantum discord and consequently the LII functions essentially quantify the difference between the conditional entropy after and previous optimized measurements, it is useful to quantify the  LII amount involved when measurements are made in a sequential closed form, \textit{e.g.} $E\rightarrow B\rightarrow A$.  In that sequence the LII in the pure tripartite system $ABE$, is computed by adding the pairwise QD contributions when measurements are performed on $E$, $B$ and $A$ to infer the mutual information of the pair  $EB$,  $BA$, and  $AE$, respectively,
\begin{equation}
{\cal{L}}_{\circlearrowright}\equiv \delta^{\leftarrow}_{BE}+\delta^{\leftarrow}_{AB}+\delta^{\leftarrow}_{EA}.
\label{clock}
\end{equation}
The resulting amount represents, as shown in Fig. (\ref{flow}), a clockwise, ${\cal{L}}_{\circlearrowright}$, flow of pairwise LII \cite{flowDVicHay}, and it represents how much the joint $ABE$ system state is affected by the sequential optimized measurements on $E$, $B$, and $A$.
Reversely, the computation of the pairwise QDs for the sequence of measurements $A\rightarrow B\rightarrow E$ represents a counterclockwise (see Fig.~(\ref{flow})), ${\cal{L}}_{\circlearrowleft}$, pairwise flow of LII,
\begin{equation}
{\cal{L}}_{\circlearrowleft}\equiv\delta^{\leftarrow}_{BA}+\delta^{\leftarrow}_{EB}+\delta^{\leftarrow}_{AE}
\label{countclock}.
\end{equation}

Through Eq. (\ref{Eab2}), we can extend Eqs. (\ref{law1}-\ref{law3}) to see that, for an arbitrary pure tripartite quantum system, the sum of all bipartite EOF is equal to the sum of all average LII,
\begin{equation}
E_{AB} + E_{AE} + E_{BE} = \varpi^+_{A|B} + \varpi^+_{A|E} + \varpi^+_{B|E},\label{sum}
\end{equation}
or \begin{eqnarray}
E_{AB} + E_{AE} + E_{BE} &=& \frac12\left({\cal{L}}_{\circlearrowright}+{\cal{L}}_{\circlearrowleft}\right).\label{sum2}
\end{eqnarray}
 So the sum of all possible EOF between pairs $A$, $B$ and $E$ is the sum of the clockwise and counterclockwise flow of LII. But the difference between (\ref{clock}) and (\ref{countclock}) LII flows gives
\begin{equation}\frac{{{\cal{L}}_{\circlearrowright}-{\cal{L}}_{\circlearrowleft}}}{2}=
\left(E_{AB} -\delta^{\leftarrow}_{AE}\right)+ \left(E_{AE} -\delta^{\leftarrow}_{EB}\right)+ \left(E_{BE}-\delta^{\leftarrow}_{BA} \right).\label{diff}\end{equation}
 Interestingly, the right hand side of Eq.~(\ref{diff}) is equal to the sum $S_{A|E}+S_{E|B}+S_{B|A}$, which vanishes for all pure $ABE$ joint state  \cite{power,koashi2004}. So, for pure states, $ {\cal{L}}_{\circlearrowright}={\cal{L}}_{\circlearrowleft}$
and Eq.~(\ref{sum2}) results in
\begin{eqnarray}
E_{AB} + E_{AE} + E_{BE} = {{\cal{L}}_{\circlearrowright}}.\label{sum3}\end{eqnarray}
Therefore, \textit{for a given tripartite pure state $\rho_{ABE}$ the sum of the pairwise EOF between  $A$, $B$, and $E$ is simply given by the LII flow in a closed cycle.} The implication of  ${\cal{L}}_{\circlearrowright}={\cal{L}}_{\circlearrowleft}$ in terms of the LII balance is that
\begin{eqnarray}
\omega^-_{A|B}+\omega^-_{B|E}+\omega^-_{E|A}=0,\label{eqzero}
\end{eqnarray}
i.e., all the cyclic sum of the LII balance ($E\rightarrow B\rightarrow A\rightarrow E$ or $E\rightarrow A\rightarrow B\rightarrow E$) vanish. This is simply a feature of the purity of the system - since the system is closed there is no LII missing, and so the balance is null.  In other terms, the amount of information contained in the cyclic sum of the conditional entropies balance is not disturbed by local measurements. As discussed bellow this result is the basis to obtain the most fundamental expressions relating the entanglement of formation and discord. Furthermore, it gives a very simple relation between the conditional entropy and QD.


\section{Difference between Entanglement of Formation and Quantum Discord}
An intriguing aspect is the difference between entanglement and quantum correlation. Once QD can be different from zero for separable states, it is usually assumed that it could include extra quantum correlations when compared to entanglement. For example, for a typically separable state of the form\begin{equation}
\rho_{AB}=\sum_i p_i \rho_A^i \rho_B^i,
\end{equation}
while $E_{AB}=0$, the QD vanishes if, and only if, the set of states $\{\rho_B^i\}$ is a set of orthogonal projectors (with the measurements in B).  
On the other hand, for mixed entangled states, there are some situations where the QD is smaller than the EOF. Thus, a fundamental question emerges: what in fact measures the difference between them?
By using Eq.~(\ref{Eab2}), we can write it as the difference between the EOF and the QD as:
\begin{equation}
E_{AB} - \delta^{\leftarrow}_{AB} = \varpi^-_{B|A} + \varpi^-_{A|E} + \varpi^-_{B|E},\label{dif}
\end{equation}
and so exclusively in terms of the LII balance. Notice that $E_{AB} - \delta^\leftarrow_{AB}$ can be either larger or smaller than zero, once it depends on the efficiency of determining the locally mutual information by performing measurements on each subsystem. It is natural that depending on the quantum state $\rho_{ABE}$, the efficiency that measurements performed in $E$ in order to determine the mutual information of the pairs $AE$ and $BE$ is different from the efficiency of measurements performed in $B$ to determine the mutual information of $BE$ and $AB$ (as well as from the one where measurements on $A$ in order to determine the mutual information of $AB$ and $AE$). Thus, the difference between the EOF and the QD gives the balance of such an efficiency. Furthermore, it is interesting to note that due to Eq.~(\ref{law1}) if $E_{AB} - \delta^{\leftarrow}_{AB}$ is positive then certainly $E_{AE} - \delta^{\leftarrow}_{AE}$ is negative and vice-versa.
\begin{figure}[htbp]
\begin{center}
\includegraphics[width=.4\textwidth]{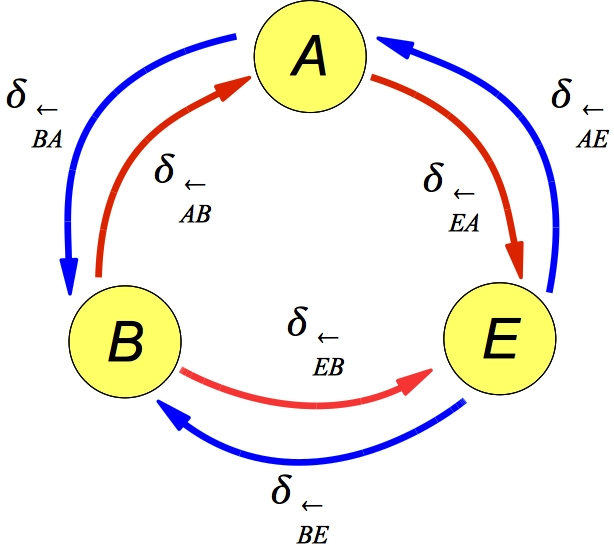} {}
\end{center}
\caption{(Color Online) Depiction of flow of Locally Inaccessible Information departing from measurements in $E$  (blue arrows) and concentrating in $E$ (red arrows). The net result of these two flows is the difference between the EOF and the QD for $AB$ when measurements are made on $B$. }\label{flow2}
\end{figure}

\begin{figure}[tbp]
\begin{center}
\includegraphics[width=.4\textwidth]{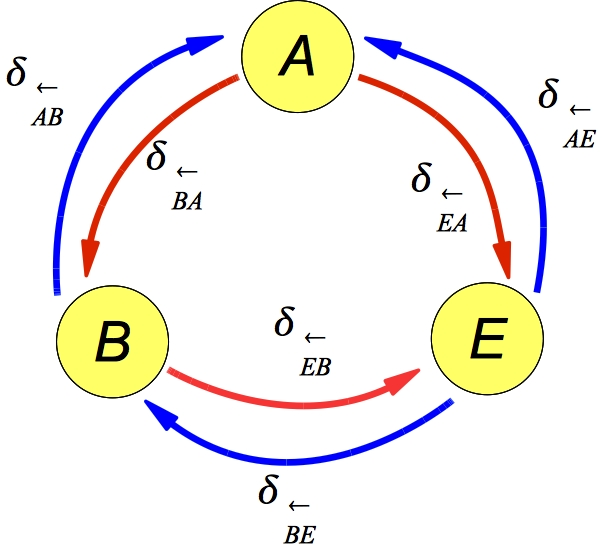} {}
\end{center}
\caption{(Color Online) Depiction of flow of Locally Inaccessible Information departing from measurements in $E$  (blue arrows) and concentrating in $E$ (red arrows). The net result of these two flows is the difference between the EOF and the QD for $AB$ when measurements are made on $A$. }\label{flow3}
\end{figure}

Eq.~(\ref{dif}) tells that the EOF and the QD differ by the amount of the LII balance in the system, but yet does not offer a clear meaning to it. This issue can be however clarified by the flow of LII as follows.
As depicted in Fig. (\ref{flow2}), all pairwise LII contributions in Eq. (\ref{dif}) can be split in two forms.
The first one is the sum of the QD with a initial measurement on $E$ and, subsequently, on $A$ concentrating on $B$, $\delta^{\leftarrow}_{AE}+\delta^{\leftarrow}_{BA}$, and with a measurements on $E$ concentrating in $B$ directly, $\delta^{\leftarrow}_{BE}$.
Similarly to what we have developed previously we can define a LII  flow  from $E$ to $B$ as
 \begin{equation}
 {\cal{L}}_{E\rightarrow A\rightarrow B}\equiv\delta^{\leftarrow}_{BE}+\delta^{\leftarrow}_{AE}+\delta^{\leftarrow}_{BA}.
 \label{LIIfromE}
 \end{equation}
 The second one  accounts for the inverse flow of LII, \textit{i.e.}, the sum of the QDs with a initial measurements on $B$ and, subsequently,  on $A$  concentrating on $E$,  $\delta^{\leftarrow}_{AB}+\delta^{\leftarrow}_{EA}$, and with a measurement on $B$  concentrating on $E$ directly, $\delta^{\leftarrow}_{EB}$. Similarly to Eq.  (\ref{LIIfromE}), we define the flow from $B$ to $E$ as
\begin{equation}
{\cal{L}}_{B\rightarrow A\rightarrow E}\equiv\delta^{\leftarrow}_{EB}+\delta^{\leftarrow}_{AB}+\delta^{\leftarrow}_{EA}.
\label{LIIfromB}
\end{equation}
Note that the definitions in Eq. (\ref{LIIfromE}) and Eq. (\ref{LIIfromB}) are asymmetric and so quite distinct from the cyclic LII flux given in Eq. (\ref{clock}) and Eq. (\ref{countclock}). With that it is possible to write Eq. (\ref{dif}) as
\begin{equation}
E_{AB} - \delta^{\leftarrow}_{AB} = \frac12\left({\cal{L}}_{E\rightarrow A\rightarrow B}- {\cal{L}}_{B\rightarrow A\rightarrow E}\right),\label{dif3}
\end{equation}
\emph{i.e.}, \textit{the difference on the entanglement of formation and the QD for the pair $A$ and $B$ when measurements are made in $B$, is the difference between the flow of LII from and to the purifying ancilla $E$. This difference is the net, or residual, LII shared with $E$.} Similarly, we can write
\begin{equation}
E_{AB} - \delta^{\leftarrow}_{BA} = \frac12\left({\cal{L}}_{E\rightarrow B\rightarrow A}- {\cal{L}}_{A\rightarrow B\rightarrow E}\right),\end{equation}
where the order of $A$ and $B$ has been changed to explicitly differ it from Eq. (\ref{dif3}) due to the distinct sequence of measurements, as depicted in Fig. (\ref{flow3}).
Combining these last two equations, it is easy to rewrite a symmetrized form for them, which turns out to be an equivalent version of Eq. (\ref{Eab2}) as
 \begin{equation}
E_{AB}-\varpi^+_{A|B}=\frac12\left({\cal{L}}_{E\rightarrow {A\choose{B}}}- {\cal{L}}_{{A\choose{B}}\rightarrow E}\right),\label{Eab3}
\end{equation}
where\begin{equation}
{\cal{L}}_{E\rightarrow {A\choose{B}}}\equiv\delta^{\leftarrow}_{AE}+\delta^{\leftarrow}_{BE},\end{equation} and
\begin{equation}
{\cal{L}}_{{A\choose{B}}\rightarrow E}\equiv\delta^{\leftarrow}_{EA}+\delta^{\leftarrow}_{EB}.\end{equation}
The form of Eq. (\ref{Eab3}) is appropriate since it is a symmetric accounting for the difference between the EOF and the average LII for the pair $AB$.
Entanglement, as measured by the EOF, is a typical correlation of quantum nature as well as the average LII, .i .e,  the average amount of mutual information locally inaccessible by measurements on $A$ and $B$. Now the difference between these quantities for  the pair $AB$ is equal to the net flow of LII between in and out of the ancilla $E$. Since ${\cal{L}}_{E\rightarrow {A\choose{B}}}$ accounts for how much the state $\rho_{AB}$ is disturbed by measurements on the purifying ancilla $E$ and, similarly,   ${\cal{L}}_{{A\choose{B}}\rightarrow E}$ accounts for how much the state $\rho_{E}$ is disturbed by local measurements on $A$ and $B$, the net LII flux ${\cal{L}}_{E\rightarrow {A\choose{B}}}- {\cal{L}}_{{A\choose{B}}\rightarrow E}$ computes the asymmetry in this process. In fact, the asymmetry captures the notion that some extra local inaccessible information of the pair $AB$ is being shared with $E$, being the reason for the difference between $E_{AB}$ and $\varpi^+_{A|B}$.
Indeed $E_{AB}=\omega^+_{A|B}$ when the system is symmetric so that $ \delta^{\leftarrow}_{AE}=\delta^{\leftarrow}_{EA}$ and $\delta^{\leftarrow}_{BE}=\delta^{\leftarrow}_{EB}$.
But the net flow of LII in and out the ancilla $E$ can vanish as well when $ \delta^{\leftarrow}_{AE}=\delta^{\leftarrow}_{EB}$, and $\delta^{\leftarrow}_{BE}=\delta^{\leftarrow}_{EA}$.
In such a case, even though $E_{AB} \neq \delta^{\leftarrow}_{BA}\neq\delta^{\leftarrow}_{AB}$, the EOF $E_{AB}$ is equal to the average LII of the pair $AB$.
So, whenever the net flow of LII in and out the ancilla $E$ is null, even though there might be some LII for the subsystem $AB$ missing for being shared with $E$, it is compensated and so $\varpi^+_{A|B}$ computes all the LII which is useful for nonlocal tasks as entanglement of $A$ and $B$ can be.

\section{Example: LII and Entanglement Sudden Death}
\begin{figure}[htbp]
\begin{center}
\includegraphics[width=.48\textwidth]{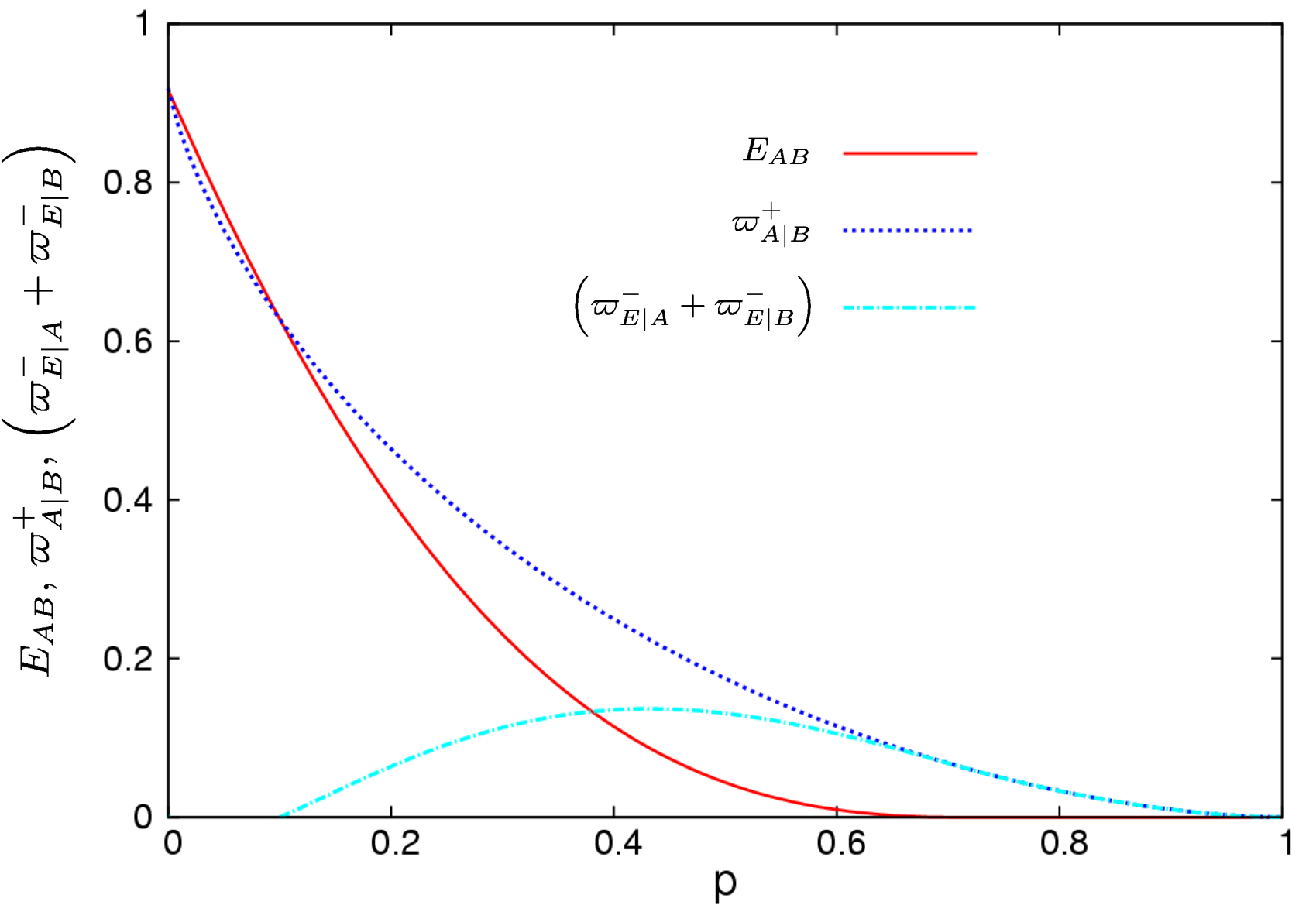} {}
\end{center}
\caption{(Color Online) the red curve (solid) shows the entanglement between the pair $AB$, while the blue curve (dotted) shows the average LII $\varpi^+_{A|B}$. The cyan curve (traced) represents the sum of the balance LII between the environment and the pair $AB$. 
When $p\approx0.65$ the average becomes equal to the sum of the balance and the entanglement sudden death occurs. }\label{fig3}
\end{figure}

By using the relations here presented, we can investigate another important aspect of the distribution of the entanglement
and the quantum discord in a multipartite system. We consider a four qubit system where two initially pure entangled qubits $A$ and $B$ interact individually with their own reservoir $R_A$ and $R_B$, respectively (for details see \cite{retamal}). We suppose an amplitude damping channel at temperature $T=0$ K and we write a map to each qubit as
\begin{eqnarray}
\hspace{-0.7cm}\Sigma\left(|0\rangle_{A}|{0}\rangle_{R_{A}}\right) & \rightarrow & |0\rangle_{A}|{0}\rangle_{R_{A}}\nonumber\\
\hspace{-0.7cm}\Sigma\left(|1\rangle_{A}|{0}\rangle_{R_{A}}\right) & \rightarrow & \sqrt{1-p}|1\rangle_{A}|{0}\rangle_{R_{A}}+\sqrt{p}|0\rangle_A|{1}\rangle_{R_{A}},\label{dyn}
\end{eqnarray}
where $p=1-e^{-\Gamma t}$ and identically for $B$ interacting with $R_B$. We choose as the initial condition $|\Psi(0)\rangle = \frac{2}{\sqrt{3}}|0\rangle_A|0\rangle_B+\frac{1}{\sqrt{3}}|1\rangle_A|1\rangle_B$, which is an example where the phenomenon known as entanglement sudden death \cite{ESD} occurs. As one can observe by Eq.~(\ref{Eab2}), the entanglement between $AB$ suddenly vanishes when the average LII between $AB$ is equal to the balance of the LII between the environment and each subsystem ($A$ and $B$). Actually, as soon as measurements over the environment allows more inference about the mutual information with the pair $A,B$, their entanglement decreases. As illustrated in Fig. (\ref{fig3}), when the entanglement between $A$ and $B$ vanishes (entanglement sudden death), the excess of the knowledge that the environment $E$ has about the subsystem $A$ and $B$, as measured by
\[
\varpi^-_{{R_A}{R_B}|A} + \varpi^-_{{R_A}{R_B}|B}={\cal{L}}_{{R_A}{R_B}\rightarrow {A\choose{B}}}- {\cal{L}}_{{A\choose{B}}\rightarrow {R_A}{R_B}},
\]
 becomes equal to the average LII in a finite time.

 {To obtain the results plotted in Fig.~(\ref{fig3}), we analytically solve the dynamics of $E_{AB}$ and the QD between each subsystem $A$ and $B$ with the whole environment $E\equiv R_A\otimes R_B$. In this case, we use our relations to analytically calculate the QD for a system of dimension $2\times4$.
For example, to calculate the QD between $A$ and the whole environment $R_A\otimes R_B$ we have that
\begin{equation}
\delta^\leftarrow_{A({R_A}{R_B})}= E_{AB}+ S_{A|B},
\end{equation}
 where $S_{A|B}$ is the conditional entropy and both, $E_{AB}$ and $S_{A|B}$, can be calculated analytically by means of the density matrix $\rho_{AB}$.
  These results extend further the investigation from Ref.~\cite{maziero} as it provides a way to calculate the QD and the EOF for different partitions (e. g. $A(R_A R_B)$) and for higher dimensional systems. Indeed, the monogamic relation can be used to calculate the QD and the EOF between two subsystems with dimension $2\times N$ and rank 2 (see also \cite{retamal2}). It is true because the extra system that purify a rank 2 density matrix is always a qubit - For example, given a qubit $A$ and an environment $E$ with dimension $N$, a rank 2 density matrix $\rho_{AE}$ can be purified in a density matrix $\rho_{ABE}$ where the dimension of the subsystem $B$ is always two. Noting that $AB$ is thus a system composed by two qubits, we have \cite{power}
\begin{equation}
E_{AE}= \delta_{AB}^\leftarrow + S_{A|B},\label{monog}
\end{equation}
and
\begin{equation}
\delta^\leftarrow_{AE}= E_{AB}+ S_{A|B}.\label{monog2}
\end{equation}
Eq. (\ref{monog}) shows that the EOF between a qubit and a qudit for any rank 2 density matrix can be calculated numerically by means of the QD of the two qubits $AB$. More importantly, Eq. (\ref{monog2}), shows that the QD between a qubit and a qudit for any rank 2 density matrix can be calculated analytically by means of the EOF of the two qubits system $AB$.}


\section{Additional Fundamental Relations}
Based upon the previous results, we are able to obtain additional fundamental expressions relating the entanglement of formation and quantum discord as well as the conditional entropy. First of all, we combine Eq.~(\ref{Eab2}) and Eq.~(\ref{eqzero}) to show that
\begin{eqnarray}
E_{AB}&=&\delta_{AB}^\leftarrow + \delta_{BE}^\leftarrow - \delta_{EB}^\leftarrow,\label{minimal}\\
E_{AB}&=&\delta_{BA}^\leftarrow + \delta_{AE}^\leftarrow - \delta_{EA}^\leftarrow.\label{minimal2}
\end{eqnarray}
These equations are the simplest expressions relating EOF exclusively to QD. They show that the difference between the entanglement and the quantum discord is proportional (twice) to the LII balance of one of the subsystem with the environment.
In Eq.~(\ref{minimal}), we see that for an arbitrarily mixed system $AB$, when the subsystem $B$ is measured in order to know about the mutual information of the pair $AB$, some additional information is acquired about the pair $BE$, and so it needs to be taken in account ($2\omega_{B|E}^-$ needs to be summed). We also can derive explicitly an important result discussed in the previous section but in a simpler form:  \textit{if the systems $B$ and $E$ are symmetric, we have that $E_{AB}=\delta_{AB}^\leftarrow$}, which is a direct consequence of the fact that for this case $\delta_{BE}^\leftarrow =\delta_{EB}^\leftarrow$.
Then, the quantum discord $\delta_{AB}^\leftarrow$ is equal to the entanglement of formation $E_{AB}$ not only when the system $AB$ is pure but also when the systems $E$ and $B$ are symmetric. The same is valid for Eq.~(\ref{minimal2}). The $\delta_{AB}^\leftarrow$ is equal to $E_{AB}$ not only when the system $AB$ is pure but also when the systems $E$ and $A$ are symmetric. Of course, if we have full symmetry between $A$, $B$, and $E$ then $E_{AB}=\omega^+_{A|B}$.

By using the results above, we are able to find a very useful relation between the quantum discord and the conditional entropy. The conditional entropy is an important quantity in information theory that is intimately related to the entanglement distillation and irreversibility. In addition, it is fundamental in the state merging protocol  \cite{horo}. In this protocol, given two parties $A$ and $B$ with a shared state $\rho_{ab}$, the conditional entropy measures the amount of quantum communication that is needed to
transfer the part  $A$ to the part $B$ such that part $B$ ends with the state $\rho_{ab}$, keeping possible correlations of $\rho_{ab}$ with any external system.
Interestingly, the conditional entropy can be negative and
this means that $B$ can obtain the full state $AB$ using only classical communication. Additionally $A$ and $B$ will be able to transfer quantum information in the future at no further cost \cite{horo}.
First of all, let us pay attention to the case of pure states. For a pure state, we can write the conditional entropy as
\begin{equation}
-S_{A|B}=\delta_{BA}^\leftarrow.
\end{equation}
As exposed above, $\delta_{BA}^\leftarrow$ measures the amount of mutual information of $AB$ inaccessible by measurements on $A$. Clearly, $A$ can not transfer this information to the subsystem $B$ and consequently it is preserved for a future communication. In this sense, what could we say about mixed states? To calculate the conditional entropy for mixed states, in terms of the LII, we use the relation \cite{power,koashi2004}
$E_{AB} = \delta_{BE}^\leftarrow + S_{B|E}$ and Eq. (\ref{minimal}). Based on this equation, it is simple to show that
\begin{equation}
-S_{A|B}=\delta_{BA}^\leftarrow - \delta_{EA}^\leftarrow.\label{entropia}
\end{equation}
It is clear now, observing equation Eq.~ (\ref{entropia}), what in fact happens for mixed states. As one can see, there is an amount of LII that $A$ shares with $E$ once it can not be sent to $B$. Furthermore, since this information mutually belongs to $A$ and $E$, it can not be used jointly with $B$ for further tasks. In fact, it has to be subtracted from $\delta_{BA}^\leftarrow$.
Moreover, by using Eq.~\ref{entropia}, it is easy to analyze the negativity of the conditional entropy, which depends on the balance of LII. The sign of the conditional entropy has an important meaning in important tasks like quantum state merging and entanglement distillation. Again, as for the EOF relations, more important than the amount of LII it is the balance of LII between the subsystems $A$ and $B$ and the purifying ancilla (environment). By using the balance of LII, we are able to identify the signal of the conditional entropy. For instance, if the subsystem $A$ shares the same amount of LII (independently of the amount) with $B$ and $E$, the conditional entropy is null, $S_{A|B}=0$. If $A$ LII with $B$ is larger that the LII with $E$, the conditional entropy is negative. Obviously, $S_{A|B}$ is positive when $A$ LII with $E$ is larger than the LII with $B$.

\section{Summary}
To conclude, we presented alternative forms to interpret the Entanglement of Formation in terms of the Locally Inaccessible Information functions. Our relations based on average LII and LII balance demonstrate that the EOF can be understood for a general quantum system exclusively as a function of the LII being shared. The concept behind LII flow when sequential measurements are made is an interesting one to understand the meaning of the correlation distribution when measurements are performed. In that sense not only the way a quantum system is affected by local measurements can be quantified but also the symmetry of such a system, under local measurements. An example of the usefulness of these new relations was given in the investigation of the yet not well understood  entanglement sudden death phenomenon. Also a deep discussion on the distinction between Entanglement of Formation and Quantum Discord in terms of residual flow of LII in and out a purifying ancilla is made possible. The relation of the QD to the conditional entropy is quite important for the understanding when the negative signal of the conditional entropy occurs.  That is ruled by the QD balance between the environment and the system as well. 
We believe the discussion presented here may contribute further for the understanding of distribution of entanglement and quantum correlation in general for multipartite systems.

This work was supported by FAPESP and CNPq through the
National Institute for Science and Technology of Quantum
Information (INCT-IQ). MCO acknowledges support by iCORE.


\begin{thebibliography}{99}
\bibitem{footnote1} Optimal measurements are those that account for base state projections that minimize $S(\rho_A|\rho_B)$.
\end{thebibliography}
\end{document}